\documentclass[sn-aps, iicol]{sn-jnl}

\jyear{2022}%

\theoremstyle{thmstyleone}%
\theoremstyle{thmstyletwo}%

\theoremstyle{thmstylethree}%

\begin{document}

\title[Article Title]{A Digital Twin-based Smart Home: A Proof of Concept Study}


\author[1]{\fnm{Laura} \sur{Bragante Corssac}}\email{lbrcorssac@inf.ufrgs.com}

\author[1]{\fnm{Juliano} \sur{Araujo Wickboldt}}\email{jwickbolt@inf.ufrgs.br}

\affil[1]{\orgdiv{Instituto de Informática}, \orgname{Universidade Federal do Rio Grande do Sul (UFRGS)}, \orgaddress{\street{Bento Gonçalves}, \city{Porto Alegre}, \postcode{91509900}, \state{Rio Grande do Sul}, \country{Brazil}}}


\abstract{A Digital Twin is a virtual system that can fully describe a physical one. It constantly receives data from its counterpart's sensors, consults external sources, and obtains manual inputs from its stakeholders. The DT uses all this information to make various computations, such as analyses, predictions, and simulations, and then possibly sends the results back to the physical system. Domotics, or Smart Home Technologies, brings intelligence and comfort to a house by automating some of its functions. Although the research on both themes is vast, there are few implementations of Smart Homes based on Digital Twin technologies, and this work aims to prove that residences can also benefit from this concept. We implement two different use cases showing that a two-way connection between a home and its virtual counterpart can provide its owners with analyses and simulation-based automation. The first study case allows the user to visualize their home appliances' past and present state concerning their energy consumption. Based on heating simulations, the second determines the best time to turn on a heater to increase the home's thermal comfort and reduce energy usage.}




\maketitle

\section*{Declarations}

\textbf{Conflicts of interests: }The authors declare no competing interests.

\section{Introduction}\label{sec:introduction}

A Digital Twin (DT) is a virtual replica of a physical system bidirectionally connected to it \cite{grieves:2017}. All data from the physical environment, collected mainly by sensors, transit in real-time to the DT, which uses this information to play its core roles: the predictive and the interrogative. The first is about predicting the future state of the physical system, including anomaly detection, fault prevention, or the prediction of a given property (e.g., temperature or pressure) of an object. The virtual replica may also make some decisions from its computations and send instructions back to the real world, such as firing an alarm in case of danger or turning off a device after noticing its faulty operation. In turn, the interrogative role is about displaying the current and past state of the physical system. The user could monitor the sensors' readings and ask questions regarding so.

Another trending subject, which helps the DT in its real-time data collection, is the Internet of Things (IoT). It proposes that computers sense the physical world autonomously by adding sensors, electronics, devices, and internet connectivity to objects, i. e., \textit{things} \cite{NEGRI:2017} \cite{ashton:2009}. IoT can act in a variety of fields. For example, nowadays' watches equipped with heart beat sensors can improve clinical diagnosis. Another niche is Domotics, or Home Automation, which involves controlling and monitoring home devices, called Smart Home, in a unified system \cite{Miori:2014}. IoT technologies can, for example, help make homes more secure \cite{keat:2018} and more energetically efficient \cite{Motlagh:2018}.

The Digital Twin is increasingly gaining the attention of academia and industry. However, there is not much research on DTs for residential places to provide their inhabitants with more convenience in daily activities. With this in mind, this work aims to implement a proof of concept DT of a residence. It aims to make a house more pleasant to its owners with a different approach than traditional Smart Home technologies, including simulation-based automation, event prediction, and past and present data analysis. 

We evaluate our proposal with two use cases. The first one focuses on the interrogative role of the DT and acts as a monitoring system of the energy consumption of a residence. This scenario gives people more information about their habits, enabling them to rationalize their power usage. The second, in turn, focuses on the predictive role of the concept. It decides the best time to turn on a room's heater so that its indoor temperature equals a target value at a determined time. It makes the decision consulting simulation data to predict the heating of the space and successfully provides better thermal comfort to the users. 

We organize the rest of the paper as follows: In Sec. \ref{sec:background}, we present concepts related to this project and related studies. Sec. \ref{sec:proposal} proposes the architecture of a residence's Digital Twin and enabling technologies. Sec. \ref{sec:usecases1} and \ref{sec:usecases2} explain the two use cases and their implementation. We then finish this work by exposing conclusions and future work possibilities in Sec. \ref{sec:conclusion}.

\section{Background Information}
\label{sec:background}

In this section, we first describe the main concepts that our research involves. Sec. \ref{sec:iot} explains how the Internet of Things (IoT) can bring people comfort and well-being by monitoring their activities and homes. Sec. \ref{sec:dt} presents the definition and roles of a Digital Twin (DT). In Sec. \ref{sec:dtuc}, we describe how related work is already using this technology. Since there is often an overlap between the terms IoT, DT, simulation, and Smart Home, we describe in Sec. \ref{sec:comparison} the definitions we adopted for a better understanding of our work.

\subsection{Internet of Things (IoT)}
\label{sec:iot}

The Internet of Things (IoT) proposes that objects, i. e., \textit{things}, should be connected to sensors, electronics, devices, and the Internet so that computers can sense the physical world autonomously \cite{ashton:2009} \cite{NEGRI:2017}. The domains of IoT applications, also known as IoT verticals, are divided differently in the literature. Rayes et al. \cite{Rayes:2019} propose the following categories: agriculture and farming, energy, oil and gas, enterprise, finance, healthcare, industrial, retail, and transportation. 

Under \textit{healthcare}, for example, some applications, such as fall detection, already improve the well-being of their users by monitoring their vital aspects with wearable devices, such as accelerometers \cite{Rayes:2019}. Other applications, such as the one presented by Akhbarifar et al. \cite{Akhbarifar:2020}, propose a health monitoring system that helps physicians diagnose while protecting patients' privacy. In some e-health applications, data mining, time-series pattern recognition, and machine learning can also be powerful tools for generating user health insights \cite{Farahani:2020}.

Rayes et al. \cite{Rayes:2019} cite Smart buildings and homes under \textit{energy} and \textit{enterprise} verticals. The first covers use cases that rely on the IoT's reliable and affordable sensors to optimize a residence's energy consumption while maintaining people's comfort. For instance, Motlagh et al. \cite{Motlagh:2018} present a use case for energy-saving and luminous comfort. Their system automatically controls a dimmable lamp considering the user's preferences and current ambient lighting perceived by light sensors. 

In turn, the \textit{enterprise} IoT vertical covers works that use automation to provide buildings and houses with safety and comfort. For instance, Keat et al. \cite{keat:2018} propose a home monitoring system that, with every detected motion, captures an image, saves it, and immediately notifies the user. 

\subsection{Digital Twin}
\label{sec:dt}

Who first introduced the concept of the DT was Grieves, defining it as a digital, complete, and accurate replica of an object \cite{grieves:2017}. The information from the sensors installed in the first transits to the second in real-time. Moreover, the DT receives data from external APIs and its stakeholders and uses this aggregated information to produce simulations and analyses, predict failure, and more. These computation results may give insights to the users or lead to autonomous actions, such as triggering an actuator. Grieves lists two core purposes of the DT:

\textbf{1. Predictive:} A DT can produce data-driven analyses of one physical product or a combination of them to predict the future behavior of the physical world. For example, a DT of an airplane could collect flight data in real-time to front-run a simulation and alert the pilots to make some decisions. Also, a DT can reduce the waste of resources by allowing inexpensive destructive tests.

\textbf{2. Interrogative:} The DT can allow users to inspect the current and past state of the physical system, showing them the sensors' readings. This data may be aggregated and used for further correlations, predictions, and fault detection.

NASA sees these roles of a DT as very promising for their products \cite{Glaessgen:2012}. As future vehicles evolve, they will require more reliable techniques of simulation and monitoring. An ultra-high fidelity mirror of an automobile could, among other things, signalize anomalies, produce constantly refined predictions, and help mission managers simulate the impact of in-flight changes. 

Regarding buildings, Khajavi et al. \cite{Khajavi:2019} list some of the considerable benefits of creating a Digital Twin in this sector: (1) gathering, generating, and visualizing the environment of the building; (2) analyzing data irregularities; (3) optimizing building services. The system could detect defects in a building's services and immediately notify the mechanics, reducing the inconvenience to the residents. Real state agencies, in turn, could profit from the data from a set of facilities to make better future decisions. Moreover, high-quality virtual visits to a facility can replace costly physical ones\footnote{What are digital twins in smart buildings? - \url{https://inbuildingtech.com/bms/digital-twin-commercial-office-building/} Visited on: 6 Feb. 2022.}.

Grieves \cite{grieves:2017} proposes the ideal DT as an exact mirror of a physical product, such that a human could not distinguish a camera's stream of a physical product from its DT. We are more skeptical regarding this concept, and, for the scope of this work, a DT replicates only the information it needs for its goals.

\subsection{Digital Twin Use Cases}
\label{sec:dtuc}

Liu et al. \cite{Liu:2020} propose a DT for a building as an indoor safety management system (ISMS). The virtual replica constantly receives data from sensors installed in a facility and allows users to monitor this data in a 3D web representation of the installation. Moreover, the system uses this data, alongside AI models, to detect fire risk, overcrowding, or intruders. If so, users, which may be the security personnel of the building, receive alerts in the physical and the virtual environment as sounds and texts, respectively. The authors proved the feasibility of their method by implementing it in a building of the bobsleigh and sled stadium for the 2022 Beijing Winter Olympic Games. 

Khajavi et al. \cite{Khajavi:2019} propose a method for establishing a Wireless Sensor Network (WSN) on a building facade to enable the monitoring of some of its characteristics, such as light, temperature, and humidity. As a test bed, the authors placed sensors on a section of a building's facade to measure, among other things, its illumination at different points. The DT receives the data sensed in real time and colors the wall's 3D model based on it. The more intense the color of a region of the virtual facade, the more luminous that part of the physical counterpart is. Thus, the experiment shows a DT as a real-time monitoring system. The paper provides a detailed explanation of their implementation and proposes how one could extend it. Also, the authors suggest some applications which could use the collected data to benefit the buildings, such as an air conditioning system. However, other possible applications, such as some involving predictions and more complex data analysis, are not as detailed. Moreover, the use case only shows a one-way connection between the physical and the digital systems, lacking a procedure for sending feedback to the building.

Kaewunruen et al. \cite{Kaewunruen:2019} create a building's DT to see how to transform it into a Net Zero Energy Building (NZEB). The authors first explain the concept of a Building Information Model (BIM), a fully-descriptive 3D model of a building that produces estimations related to cost, energy consumption, and maintenance management. They created one BIM for the existing building, and another was a copy with changes on some elements, such as windows and walls, to improve the thermal efficiency. They calculated that the modifications decreased energy consumption by 6,67\%. Moreover, they simulated the behavior of the hypothetical building and proved that it could be energetically self-sufficient with additional solar panels and wind turbines. The term DT in their work is a synonym of a BIM model, which is different from what we consider since it does not work with real-time data \cite{Khajavi:2019} and an autonomous connection between the virtual and the physical systems. Moreover, while a BIM's users are often designers, architects, and engineers, those of a home's DT can be with its residents. However, their proposal and results are relevant to our discussion.

Bairampalli et al. \cite{Bairampalli:2021} make a DT of their Italian coffee machine to detect anomalies in the coffee-making process. They placed temperature and pressure sensors on the object of study and built a data set by preparing coffee under different conditions while collecting data from the sensors. The final data set allowed an AI algorithm to reach some conclusions, such as the relationship between the temperature during the process and the water amount used. The work presents in detail the data collection and analysis. However, it lacks details about other components of the digital twin, such as data storage. Also, it does not show a bi-directional connection between the digital models and the Moka and user interactions.

Another sector that has its attention to the DT is manufacturing. This technology can provide the manufacturers with real-time monitoring of their machinery and production line and, thus, the ability to predict issues sooner. Zheng et al. \cite{Zheng:2019} propose an application framework of DT and built a DT for a welding production line, mirroring its geometric, physical, and kinematic properties. The connection between the virtual and physical systems allows fault detection and a remote, real-time, and visual monitoring of the production process. 

The last field of application of DT that we investigate is healthcare. Liu et al. \cite{Liu:2019} define the concept of Digital Twin Healthcare, which receives data from a patient, such as real-time heart rate measurements and blood test results, and from other environmental elements that impact a person's health, such as the weather. The system can aid in a patient's supervision by healthcare givers and provide medication reminders, crisis warnings, and medication suggestions. Also, the DT can help hospital administration by aggregating data from different patients and predicting times of high demand for medical care. 

Although the presented studies cover different roles of the DT, none of them apply the concept of a DT in a house to be used by its owners. In addition, none of the approaches demonstrates the interrogative role of the DT and allows users to interact with the system by asking questions about past data. Our work shows that a home's DT is possible and can provide its homeowners with real-time information and automation. Also, we show how to include their inputs on the proposed framework and demonstrate this interaction.

\subsection{Comparison between terms}
\label{sec:comparison}

Some confusion may arise by distinguishing DTs from general computing models and simulations \cite{Griffor:2017}. IBM states on its website\footnote{What is a DT: \url{https://www.ibm.com/topics/what-is-a-digital-twin}. Visited on: 6 Fev. 2022.} that the difference stays on the scale. While simulations usually apply to one single process, a DT studies the combination of various them. Moreover, simulations commonly do not benefit from a bi-directional data integration with the studied system. 

To conclude this section, we differentiate traditional Smart Home technologies from a home's Digital Twin. The first aims to bring house smartness mainly with automation. The latter is also a means of turning an ordinary home into a Smart Home; however, it has broader purposes. A home's DT can provide, beyond automation, fault detection and analysis. Moreover, Smart Home technologies usually provide automation with simple rules. The DT aims to deliver higher-fidelity automation by combining the house's physical characteristics with a simulation of its behavior. 

\section{An architecture for a home's DT}
\label{sec:proposal}

This section first presents a conceptual model of a residence's digital twin, based on the work proposed by Behrendt et al. \cite{Behrendt:2019}. We later describe the enabling technologies, focusing on those we used in our use cases implementation to facilitate their understanding.

\begin{figure}
    \centering
    \includegraphics[scale=0.1]{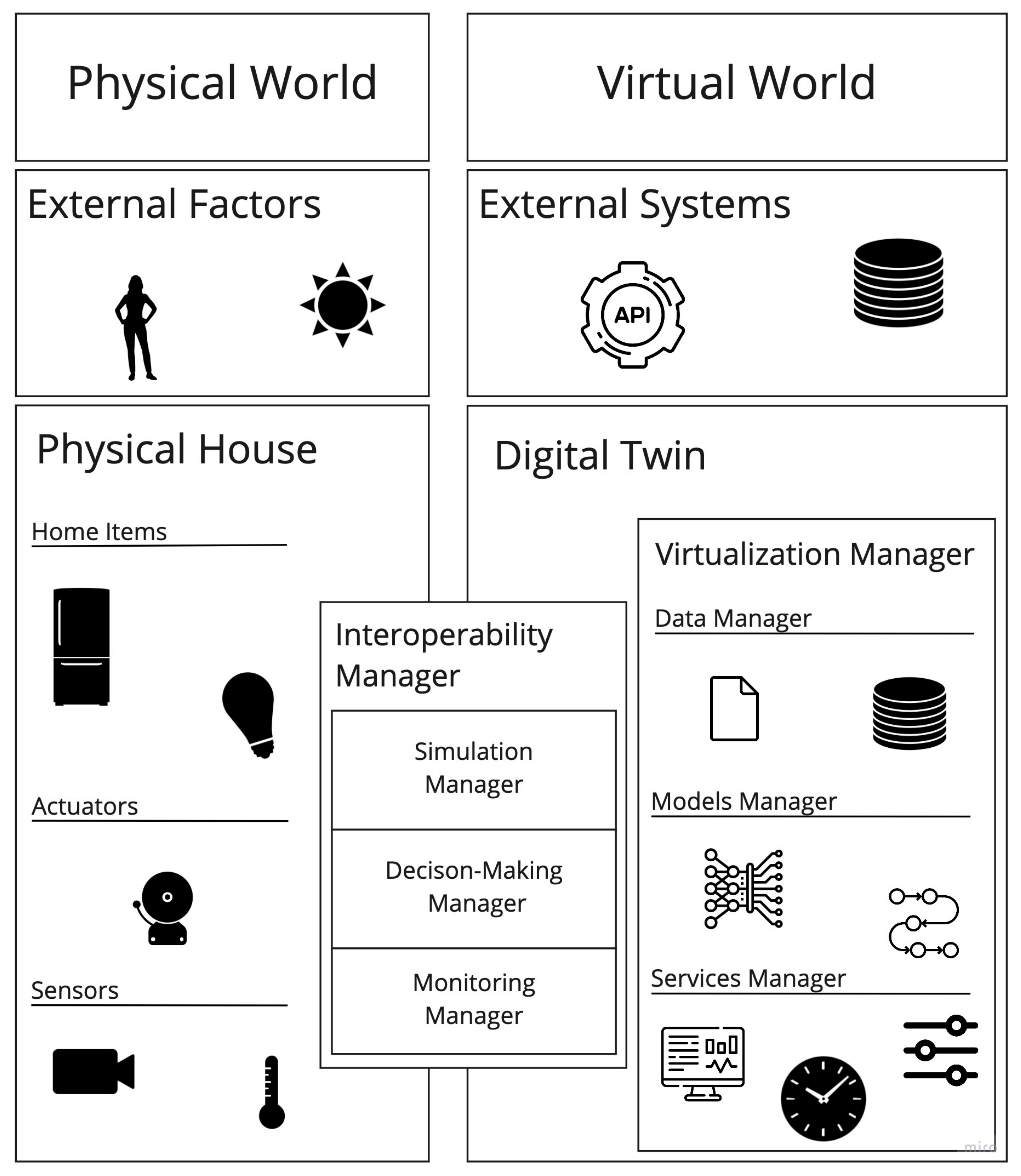}
    \caption{A Home's DT architecture}
    \label{fig:HomeDT}
\end{figure}

\subsection{Conceptual Model}

\textbf{Physical World:} It includes primarily our object of study, the house, which encompasses everyday items, such as household appliances, and sensors and actuators, which interact with the virtual world. Moreover, elements external to the house that change its state also comprise the physical world. Examples are the weather and people, including residents (i. e., the users) and visitors.

\textbf{Virtual World:} Its core component is the Digital Twin. Behrendt et al. \cite{Behrendt:2019} divide its roles into two main building blocks: the Interoperability Manager and the Virtualization Manager. The latter comprises the following three sub-components:

\begin{itemize}
    \item \textbf{Data Manager:} Its sub-component Data Acquisition Manager acquires data from different sources, including the sensors on the house, external APIs, and the system's stakeholders. The data may have diverse forms, such as a constant flow of events in the case of the sensors or non-volatile (unchangeable) data, such as the specifications of the house devices, its Bill of Materials, and its BIM. Another sub-component is the Data Analytics Manager, which stores the data and prepares it for further operations.
    \item \textbf{Models Manager:} This includes data computation models that process and analyze the data stored by the Data Manager, which may later receive and store the results. It may count on AI algorithms for different purposes, such as anomaly detection. The Model Manager also comprises data representation models, which allow data storage, exchange, and searching. It may include ontologies, taxonomies, and XML-based models.
    \item \textbf{Services Manager:} This is responsible for orchestrating the functionalities of the DT. As the DT may have different users with different interests, this module also supports a custom operation of the system. It may comprise authentication and authorization mechanisms that protect a home's information from non-allowed people. Moreover, they may allow only determined residents to use a particular feature. 
    This module also comprises visualization dashboards, which provide a graphical display of different functionalities, such as the real-time monitoring of the sensors. 
\end{itemize}

Besides the Virtualization Manager, the DT comprises the \textbf{Interoperability Manager}, which interfaces the real and the virtual domains. Its \textbf{Monitoring Manager} connects the physical sensors to the Data Manager. In turn, the \textbf{Decision-Making Manager} sends the DT's decisions to the actuators. The last sub-component, the \textbf{Simulation Manager}, provides the simulations.

\subsection{Enabling technologies}

\textbf{Data Manager:} Since IoT applications usually involve handling data from sensors, a lightweight protocol is well-suited for \textit{Data Acquisition}. We chose the connectivity protocol Message Queue Telemetry Transport (MQTT), which is popular in this context because of its push protocol aspect and bandwidth efficiency \cite{Soni:2017}. Its architecture relies on a many-to-many publish-subscribe model that comprises topics, publishers, subscribers, and a broker. Topics are ordered sequences of events. The publishers send messages to these structures, managed by the broker, which sends the information to the respective subscribers of each topic. We used this technology in the implementation of our two use cases. More information regarding it we provide in Sec. \ref{sec:uc1Monitoring}, \ref{sec:uc1Conn}, \ref{sec:uc2Monitoring}, \ref{sec:uc2DataAcq}, and \ref{sec:uc2Conn}. Other popular protocols for this purpose are the Advanced Message Queuing Protocol (AMQP) and the Constrained Application Protocol (CoAP).

Examples of free tools that may compose the \textit{Data Storage} sub-module are Apache Kafka and MongoDB, a NoSQL database. In our case, we used the first, an open-source event streaming platform mainly maintained and distributed by Confluent. It works similarly to the MQTT protocol. Its main data structure is a stream, also referred to as a topic. The \textit{Producers}, or publishers, write on topics, and the \textit{Consumers}, or subscribers, receive the events from the topics to which they subscribe. The event sequences are stored on disk, making their information accessible and queryable. We provide more practical details about this technology in Sec. \ref{sec:uc1DataManager}, and \ref{sec:uc2DataAna}. 

A link between the \textit{Data Acquisition} and the \textit{Data Storage} systems may be necessary. In our case, Kafka Connect did the work of transposing the data from MQTT topics to Kafka ones and vice versa. Its components \textit{Connectors} have two types: Source and Sink. The first, a \textit{Producer}, reads from an external service and writes on Kafka topics. The second, a \textit{Consumer}, does the opposite. Connectors for popular technologies, such as MQTT, are available on Confluent's website\footnote{Confluent Hub: \url{https://www.confluent.io/hub/} Visited on: 25 Jun. 2022}. More information about how we implemented it is available in Sec. \ref{sec:uc1DataManager}, \ref{sec:uc1Conn}, \ref{sec:uc2DataAcq}, and \ref{sec:uc2Conn}.

\textbf{Models Manager:} In our work, we used the free software Energy 2D to make a computational model of our physical house. The simulator allows us to create 2D environments and model all three modes of heat transfer - conduction, convection, and radiation. More details are present in Sec. \ref{sec:uc1CompModel}. Computational models can also apply machine learning algorithms. Examples of highly used open-source tools for this are Tensorflow, Keras, and Pytorch.

Under data representation models, we used the Simulation Description Format (SDF). As the next section describes, we used the simulation software Gazebo, which uses the latter format to represent 3D objects and environments. More information is available in Sec. \ref{sec:uc1Visualization}.

\textbf{Services Manager:} In our first use case's implementation, we used Gazebo as a visualization dashboard. The software is an open-source 3D robotics simulator that models, among other things, sensor readings and physics. We did not use for its main purpose since our application does not count on robotics. However, it fitted our implementation by allowing us to create a 3D world representing our object, a house, of study comprising 3D models ready-made or created by us to represent some of its items. Also, it let us write scripts that receive data from other parts of our system and change some characteristics of the models. More information we provide in Sec. \ref{sec:uc1Visualization}. Other examples of visualization dashboards are Kibana and Grafana. Both are open source and very popular in this matter.

To query data stored on the Data Manager component, SQL APIs may play an important role. For querying the data stored on streams in Kafka, there is a specific Java API called Kafka Streams. It allows us to fetch data from the streams and perform operations such as filtering, aggregating and joining. To this matter, our use case's services count on Kafka's ksqlDB, another API built upon Kafka Streams that enables us to query the streams with SQL-like syntax. Other SQL APIs could also play the same role in this component. We provide more information regarding it in Sec. \ref{sec:uc1Services} and \ref{sec:uc2Services}.

\section{Use Case 1 - Power consumption visualization}
\label{sec:usecases1}

This use case focuses on the interrogative role of the Digital Twin. It allows the user to monitor their home's power consumption and interact with the system to analyze past averages. It aims to give people more information about the subject since electricity is ubiquitous for many, who do not notice how much we rely on it \cite{gates:2021}. 

The idea is to constantly measure the power of some home appliances and make a virtual color-coded gradient map based on these data. The more intense the color of the virtual counterpart of a device, the more energy it is currently demanding. The users can choose the gradient color and execute some queries on the past data. For example, they may want to know the consumption average of each instrument in the last month. The DT then pauses the live procedure and assigns new colors to the appliances based on their average consumption for the period stated by the user.

\begin{figure}
    \centering
    \includegraphics[scale=0.09]{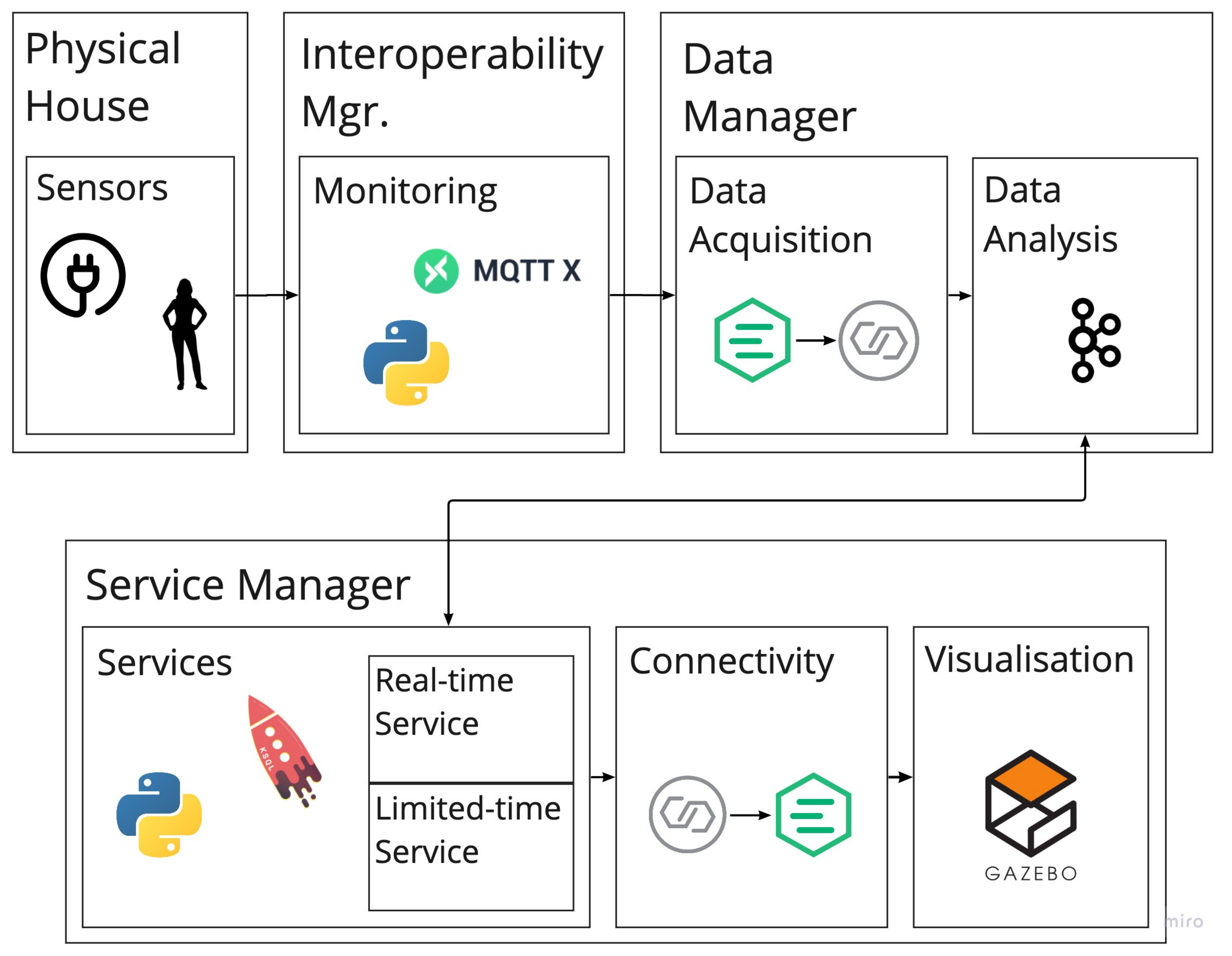}
    \caption{A DT's flow diagram}
    \label{fig:flowDiagram}
\end{figure}

Fig. \ref{fig:flowDiagram} shows the system's architecture, which relies on the model we presented in the previous section. Its building blocks we detail below.

\subsection{Physical System} 
\label{sec:uc1Physical}
It ideally consists of a house with power sensors attached to household appliances. However, to this work's scope, we simulated it with an open-source data set already compiled\footnote{\url{https://www.kaggle.com/taranvee/smart\%2Dhome\%2Ddataset\%2Dwith\%2Dweather\%2Dinformation} Visited on 15. Mar. 2022.}. It contains power readings in kW of several devices and the timestamps of the measurements, spaced by one second. For our study, we selected four machines: the microwave (Micro.), the fridge, the oven, and the dishwasher (Dish.). Table \ref{table:HomeCSV} shows a part of this CSV file with simplified columns' names and values to fit the screen. Column \textit{Time} contains how many seconds have passed since January 1, 1970 to the time when the measurement was taken. 

The users also compose the physical system, and we treat their commands like the sensors' readings.

\begin{table}[h]
\begin{center}
\caption{Sample of "Smart Home Dataset with weather information"}
\label{table:HomeCSV}
\begin{tabular}{@{}lllll@{}}
\toprule
Time & Dish. & Oven & Fridge & Micro. \\ 

(s) & (kW) & (kW) & (kW) & (kW)  \\ \midrule

1451624400 & 3.33 & 0.02 & 0.12 & 0.01 \\
1451624401 & 0 & 0.02 & 0.12 & 0.01 \\ 
1451624402 & 1.67 & 0.02 & 0.12 & 0.01\\ 
\botrule
\end{tabular}
\end{center}
\end{table}

\subsection{Monitoring}
\label{sec:uc1Monitoring} 
We built a Python script which parses the mentioned file. For each line, it creates a JSON structure for each machine, totaling four. Each machine's JSON contains three keys: (1) a string for its name; (2) a double for its power consumption; (3) a long integer for the line's timestamp. Then, the program publishes all four messages in an MQTT topic named \textit{equipment}. It waits one second and repeats the process to the next line. For example, the first line of Table \ref{table:HomeCSV} would lead to the following inputs on topic \textit{equipment}. Again we simplified the names of the machinery.

\begin{footnotesize}
\begin{verbatim}

{"equipment": "Dish.", "power": 3.33, "time": 1451624400}
{"equipment": "Oven", "power": 0.02, "time": 1451624400}
{"equipment": "Fridge", "power": 0.12, "time": 1451624400}
{"equipment": "Micro.", "power": 0.01, "time": 1451624400}

\end{verbatim}
\end{footnotesize}

Another MQTT topic named \textit{users} expects messages, also JSON-formatted, from the users. This module ideally comprises a user-friendly interface to translate the users' needs to the JSON format and publish in the user's topic. However, this did not enter the scope of this work, and we replaced it with the MQTT Client MQTT X\footnote{\url{https://mqttx.app/} Visited on 15. Mar. 2022}.

To interact with the system, the users must first download, install, and open the MQTT X platform and then start an MQTT client. If the user wants to change the color of the visualization, they must write a JSON message containing the key \textit{color} with values 1, 2, or 3, meaning red, green, and blue, respectively. If they want to see the average in a certain period, they should write a message containing: (1) the key \textit{command} with the value \textit{limited}, which tells the system to pause the real-time operation; (2) the keys \textit{from\_date} and \textit{to\_date}, indicating the initial and final dates of the period desired, respectively. Both keys' values should be a string in the format \textit{yyyy-MM-dd}. For example, if one wanted to have the average consumed by each piece of equipment in January of 2022, they would publish the following message on the \textit{users} topic: 

\begin{footnotesize}
\begin{verbatim}

{"command": "limited", "from_date": "2022-01-01",  
"to_date": "2022-01-31"}

\end{verbatim}
\end{footnotesize}

When they want to resume the real-time operation mode, they should send a message with the key \textit{command} equal \textit{realtime}.

\subsection{Data Manager} 
\label{sec:uc1DataManager}

For the data acquisition, the MQTT broker receives the messages sent by the previous component. We chose the EMQX broker because it is open source and massively scalable\footnote{\url{https://www.emqx.io/} Visited on 16. Mar. 2022}. Since the following component stores the events on Kafka topics, we must bridge the data from the MQTT topics to their respective on Kafka. To do so, we used the MQTT Connector Source available on Confluent Hub\footnote{\url{https://www.confluent.io/hub/confluentinc/kafka\%2Dconnect\%2Dmqtt} Visited on 15. Mar. 2022}. We configured it so that each of our two MQTT topics, \textit{users} and \textit{equipment}, has its destination Kafka topic with the same name. Thus, we count on two streams in Kafka that store their events for further processing.

\subsection{Services} 
\label{sec:uc1Services}

\begin{figure}
    \centering
    \includegraphics[scale=0.09]{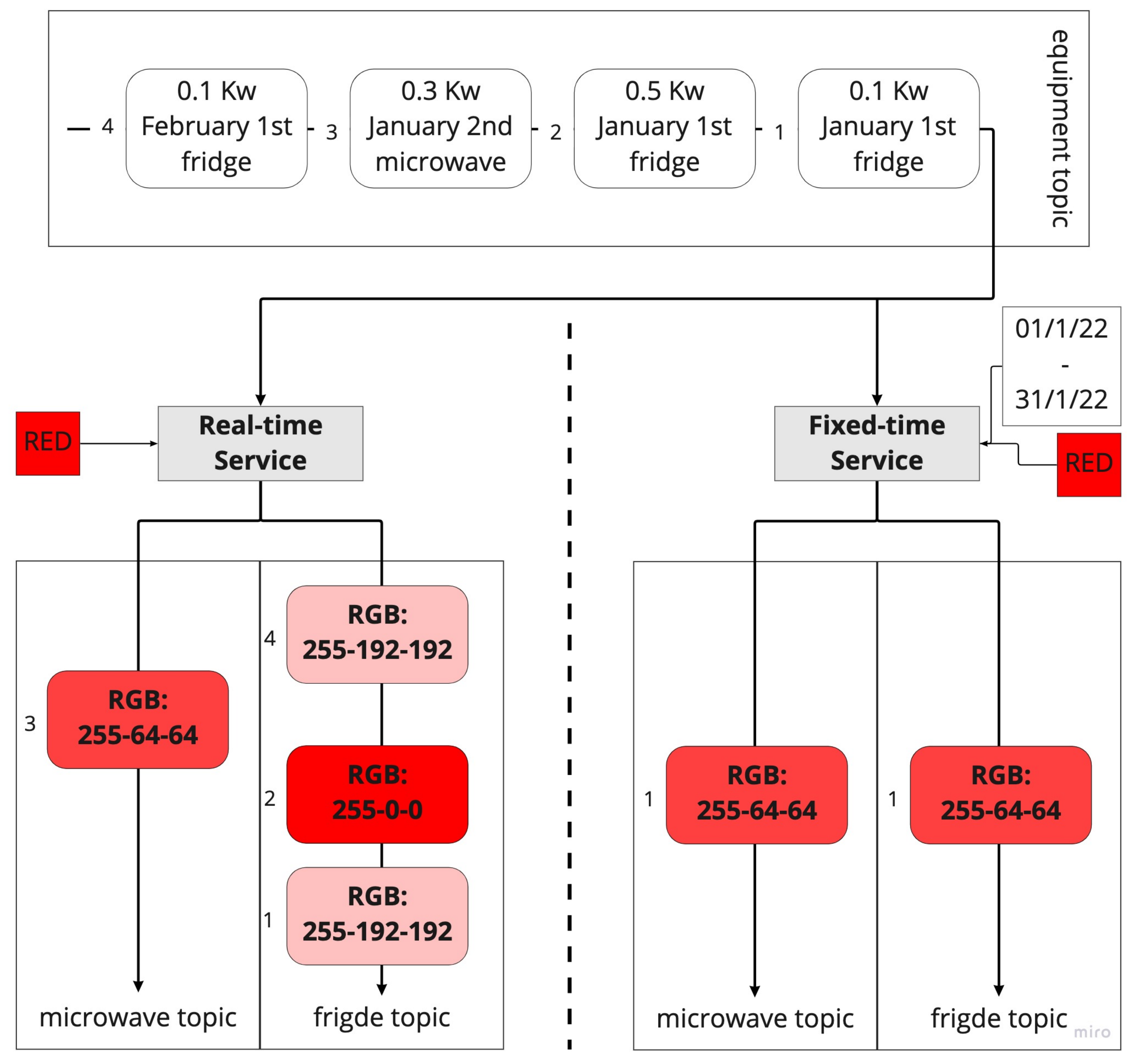}
    \caption{Service's operation modes}
    \label{fig:operationModes}
\end{figure}

We wrote a Python script to control our two services, \textit{real-time} and \textit{fixed-time}. We illustrate both operation modes in Fig. \ref{fig:operationModes} and detail them below.

\begin{figure}
    \centering
    \includegraphics[scale=0.09]{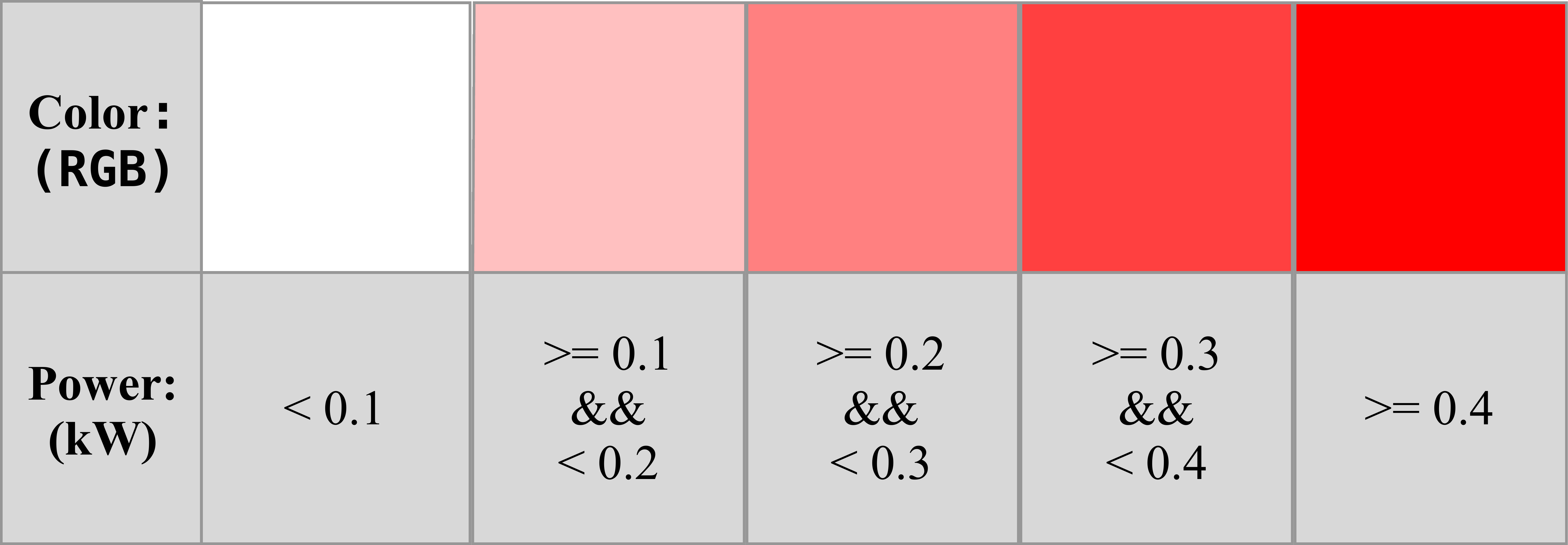}
    \caption{Association between power, in kW, and color}
    \label{fig:gradient}
\end{figure}

The system initially starts in real-time mode, listening to topic \textit{equipment}. For each message it receives, the procedure converts its power value to a color, an operation that rests on a gradient composed of 5 different shades of the last color chosen by the user, red by default. We show the mapping from power ranges to a red color gradient in Fig. \ref{fig:gradient}. The program publishes a string containing the RGB of the obtained color in a Kafka topic respective to the piece of equipment that emitted the event. Thus, we count on four output topics on Kafka named after the machines: Microwave, Fridge, Oven, and Dishwasher.

If the system receives a message from topic \textit{users} with the \textit{command} field equal \textit{limited}, it pauses the real-time processing and starts the \textit{limited-time} mode. This procedure uses the kSQL Python API to query topic \textit{equipment} and obtain the average, per device, of the power usage in the period determined by the fields \textit{from\_date} and \textit{to\_date} of the received message. For example, if the user sent the message in paragraph \textbf{Monitoring}, the following query would be performed:

\begin{verbatim}

SELECT EQUIPMENT, AVG(POWER) AS AVG_POWER
FROM EQUIPMENT_STREAM
WHERE DATE >= '2022-01-01' AND 
DATE <= '2022-01-31'
GROUP BY EQUIPMENT;

\end{verbatim}

As a result, it would obtain a list for each machine. These lists have two elements: the first contains the name of a piece of equipment, and the second its average power consumption in kW in the determined period. An example is below.

\begin{footnotesize}
\begin{verbatim}

["Fridge", 0.5], ["Micro.", 0.4], ["Dish.", 0.01],
["Oven", 0.2]
    
\end{verbatim}  
\end{footnotesize}

The algorithm then converts the four kW values to colors and publishes them on the output topics.

If a new message reaches the topic \textit{users} with \textit{command} field equal to \textit{realtime}, the program resumes coloring the machines based on their current state. If the program receives a message containing a color, independently of the operation mode, it saves this preference and considers it for following transformations.

\subsection{Connectivity} 
\label{sec:uc1Conn}
This module is responsible for connecting the two services, which work with Kafka, to the visualization component, which works with MQTT. Thus, it does a similar process as Data Acquisition, but in reverse order. We used the MQTT Connector Sink available on Confluent Hub to map our four Kafka topics to MQTT ones with the same name.

Since one of the roles of the Service Manager is to orchestrate data and provide visualization dashboards, it comprises the services, the visualization system, and the connection between them. 

\subsection{Visualization} 
\label{sec:uc1Visualization}
\begin{figure}
    \centering
    \includegraphics[scale=0.2]{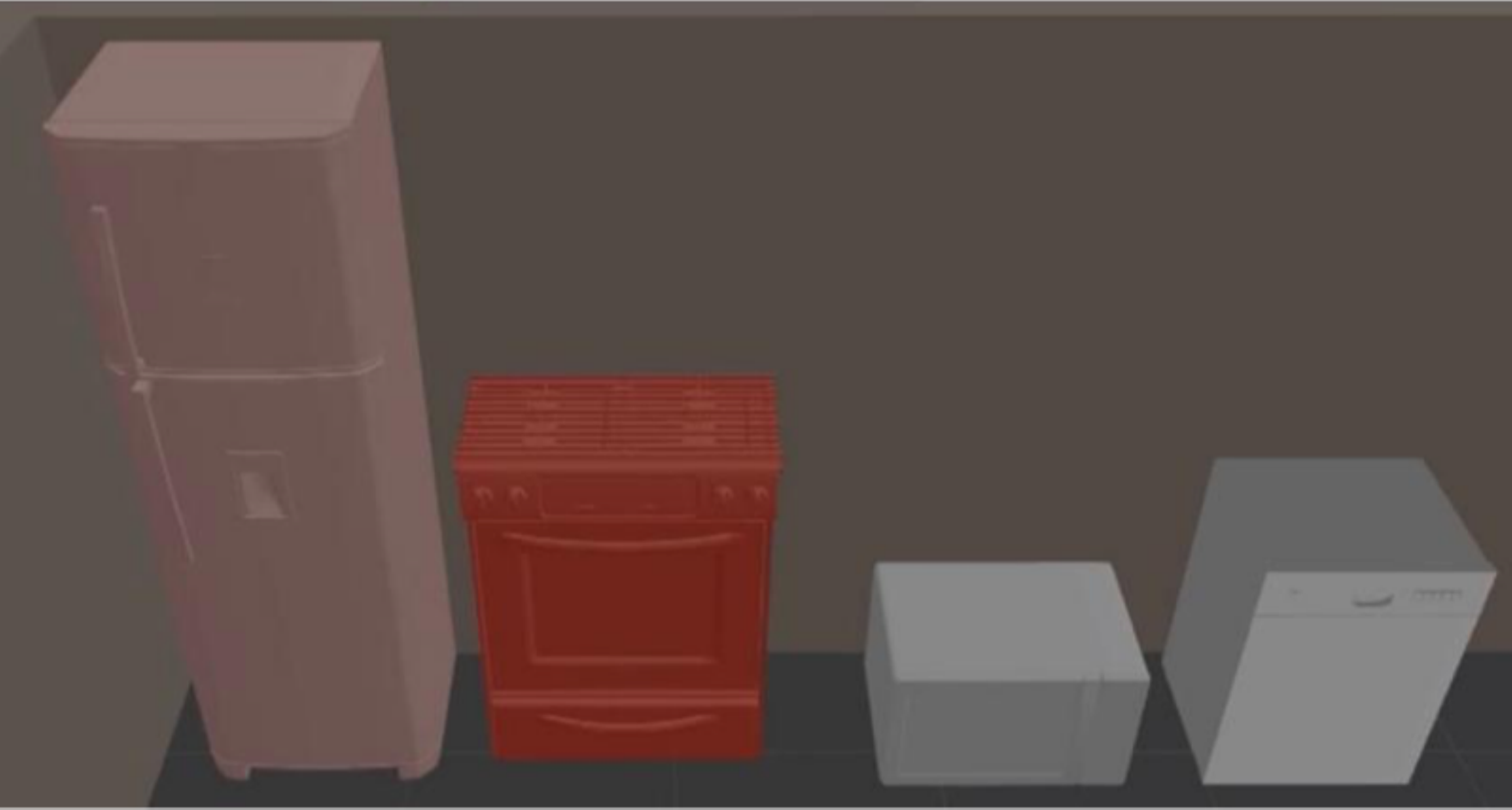}
    \caption{First use case result}
    \label{fig:UC1Result}
\end{figure}

To provide the users with a visualization of the color-coded gradient map, we chose the software Gazebo. It allowed us to create a 3D world represented by an SDF file, a format with an XML-like syntax. This file references the Collada and OBJ files of the four appliances' 3D models. We link each of them to a control script written in C++ that subscribes to the MQTT topic it obtains as an initialization argument. So, each appliance has its instance of this script. Each instance receives the name of its respective machine as a parameter. Fig. \ref{fig:UC1Result} shows the real-time functioning of the algorithm. At that moment, the oven was the most energy-consuming object; the fridge was the second; the rest had an insignificant usage.

\section{Use Case 2 - Indoor temperature control}
\label{sec:usecases2}

This use case focuses on the DT's predictive role and aims to daily obtain the right time to turn on a room's power source so that its indoor temperature equals a target at a given time. Our system constantly receives temperature measurements from the house.  Each time, it consults a simulation of the room's heating to decide if the heater should start operating. We consider that the user daily leaves home at 8 am and returns at 6 pm, the latter being the time we wish to have our target temperature of 20ºC. This section first explains the implementation of the scenario described and, in sequence, its evaluation. 

\subsection{Implementation}

\begin{figure}
    \centering
    \includegraphics[scale=0.1]{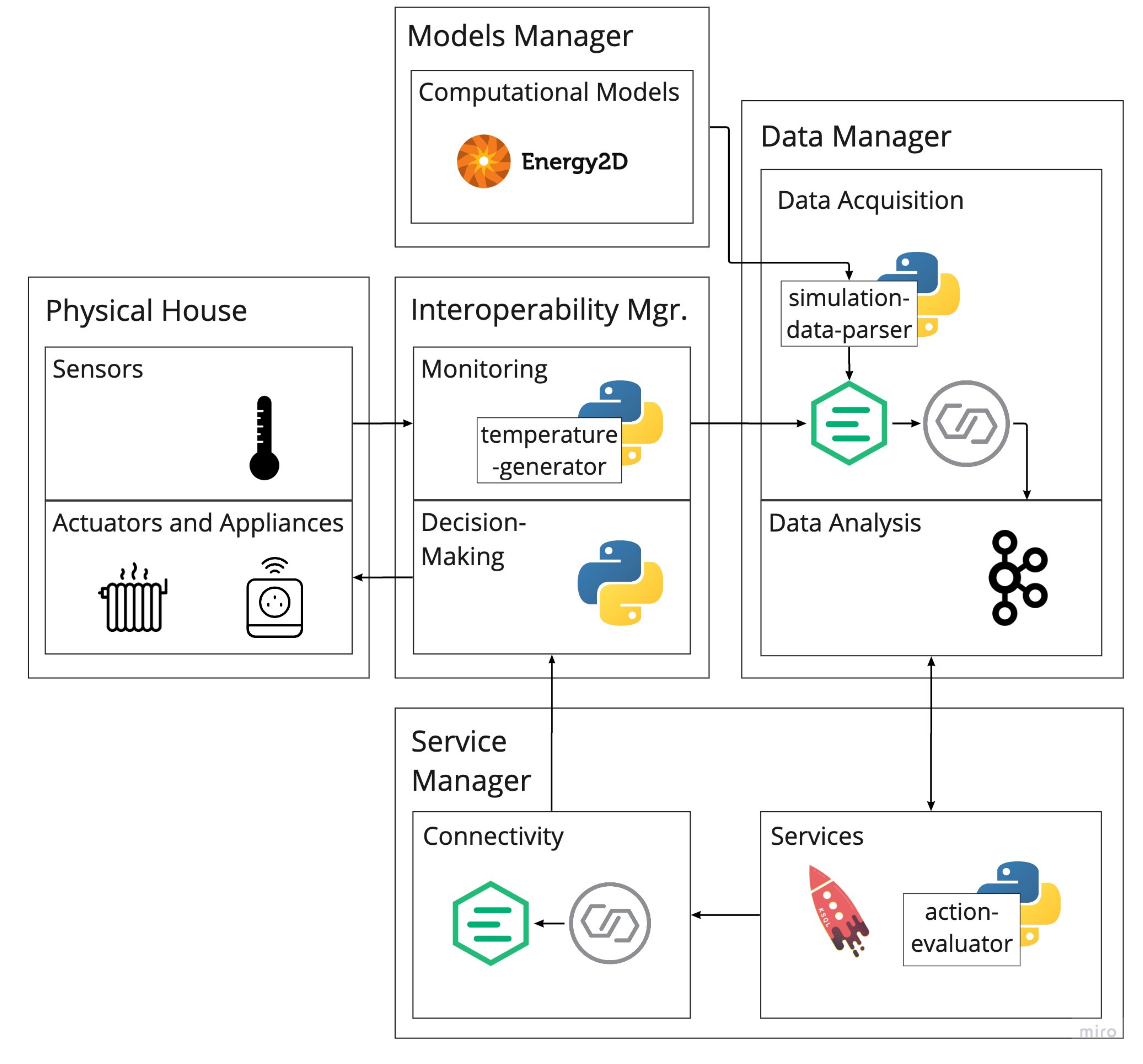}
    \caption{Second use case's architecture}
    \label{fig:usecase2Arq}
\end{figure}

We designed the second use case similar to how we did the first, taking Behrendt et al.'s \cite{Behrendt:2019} work as a reference. We show the system's architecture in Fig. \ref{fig:usecase2Arq} and describe its building blocks below. As we further explain in section \ref{sec:evaluationUC2}, we evaluate the system's operation for the first four months of the year and three different rooms. Each system's iteration considers a room and a month, information that some of our programs receive as parameters.

\subsubsection{Physical System}
\label{sec:uc2Physical}

Since we do not have an actual environment with a heater and a thermometer, we once more emulate the sensor's readings with a data set we obtained online\footnote{\url{https://energyplus.net/weather-location/europe_wmo_region_6/DEU/DEU_Stuttgart.107380_IWEC} Last visited on Jun. 17}. Table \ref{table:stuttgartDataset} shows a piece from this CSV file, which we refer to as \textit{stuttgart-weather}, and contains hourly temperature measurements in  Celsius collected over one year of the German city Stuttgart.

\begin{table}[h]
\begin{center}
\caption{Sample of \textit{stuttgart-weather} file}
\label{table:stuttgartDataset}
\begin{tabular}{{@{}lllllllll@{}}}
\toprule
month & day & hour & temperature \\ [0.5ex] 
&&& (ºC) \\
\midrule
1 & 1 & 1 & -1.9 \\ 

1 & 1 & 2 & -3.6 \\ 

1 & 1 & \multicolumn{2}{|c|}{[...]}\\

1 & 1 & 24 & -6.7 \\

1 & 2 & 1 & -6.3 \\

1 & 2 & 2 & -5.9 \\

1 & \multicolumn{3}{|c|}{[...]}\\

1 & 31 & 24 & 1.7 \\

2 & 1 & 1 & 1.5 \\
\botrule
\end{tabular}
\end{center}

\end{table}

\subsubsection{Interoperability Manager}
\label{sec:uc2Monitoring}

Has two sub-components, one to receive the sensors' readings and another to switch on the heater according to the system's conclusion. We wrote a single Python script, \textit{temperature-generator}, comprising both. It takes a month as input and selects the rows from the \textit{stuttgart-weather} file respective to that month and whose column \textit{hour} is between 8 am and 6 pm. Moreover, since our data set contains hourly measurements, we decided to provide a better precision to the whole system by generating three inter-row values following a linear progression. Lastly, the script reads the resulting file line by line, now spaced by 15 minutes. For each line, it makes a JSON message containing one key for information of hour and minute and another with the temperature at the time. Then, it publishes it in the MQTT topic \textit{temperature}. One example of these messages it publishes is the following.

\begin{footnotesize}
\begin{verbatim}

{"temperature": 19, "time": "10:15"}

\end{verbatim}
\end{footnotesize}

Moreover, the script subscribes to an MQTT topic called \textit{heater-actions}, from which it knows when it is time to turn on the heater. The program's output is a CSV file containing, for each day, the respective time at which the system decided to turn on the heater, i. e., the \textit{action-time}, and the temperature in Celsius at that moment. We refer to these output as \textit{heater-routine} files, and Table \ref{table:heaterroutine} shows a part of one.

\begin{table}[h]
\begin{center}
\caption{Sample of a \textit{heater-routine} file}
\label{table:heaterroutine}
\begin{tabular}{{@{}lll@{}}}
\toprule
date & action-time & temperature  \\
 & & (ºC) \\
\midrule 
1/Apr & 17:15 & 17.73 \\

2/Apr & 16:00 & 14.7 \\

3/Apr & 17:15 & 17.75 \\ 
\botrule
\end{tabular}
\end{center}

\end{table}

\subsubsection{Computational Models}
\label{sec:uc1CompModel}
\begin{figure}
\centering
\includegraphics[scale=0.1]{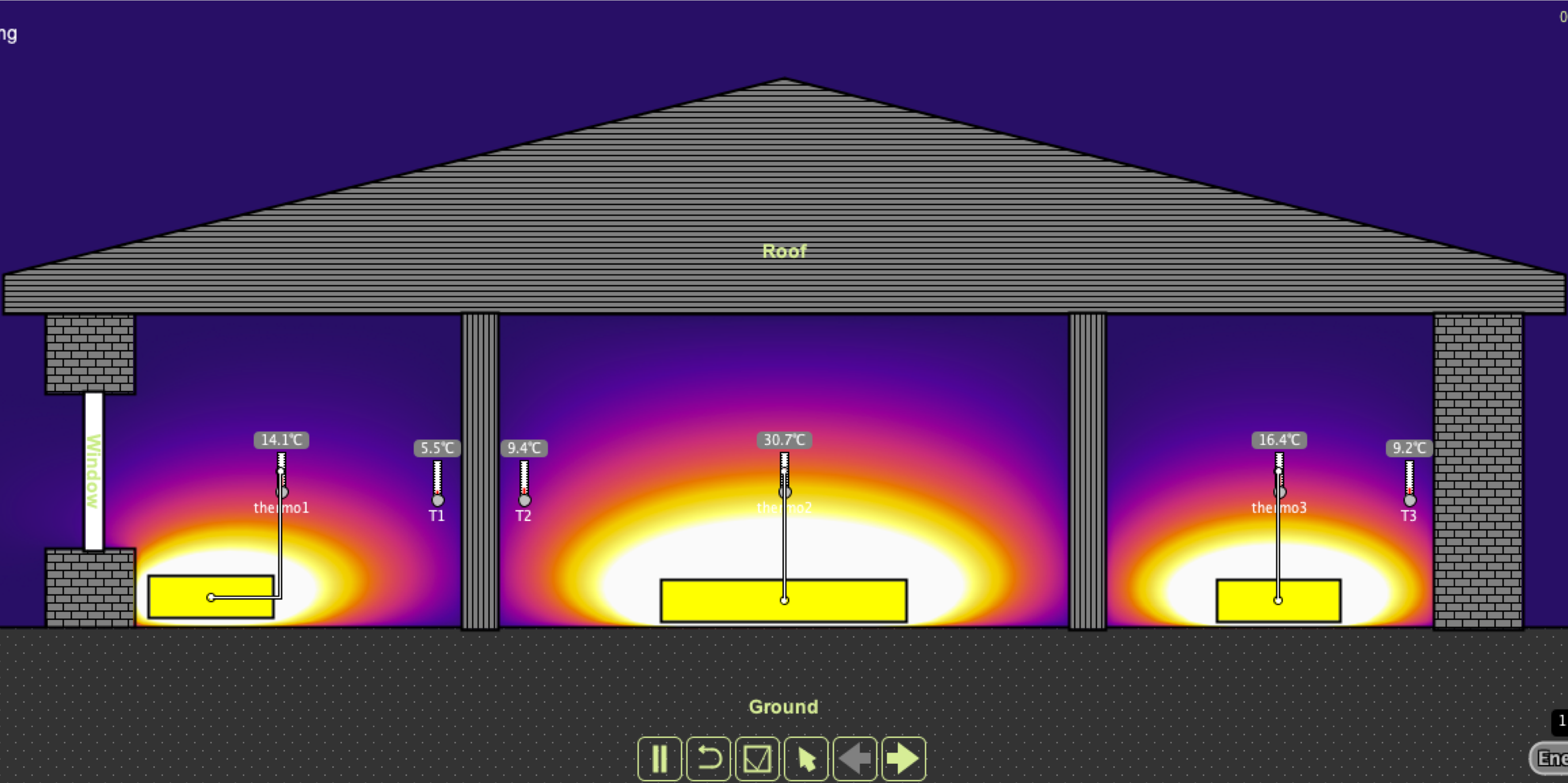}
\caption{Heating simulation running on Energy 2D}
\label{fig:simulationfinal}
\end{figure}

To model our physical house, we chose the software Energy 2D, which allowed us to create three rooms: two with smaller dimensions, one of which contains a window and the other does not. The third room is larger than the others and has no window. Fig. \ref{fig:simulationfinal} shows this configuration while the simulation was running. After the equivalent of circa 13 hours, we obtained a table containing the measurements from the six thermometers spaced by 100 seconds. We chose to consider the wall thermometers as determinants of the rooms' temperatures and ignore the columns respective to the thermostats. Table \ref{table:energy2DData} shows a part of this sub-set of columns.

\begin{table}[h]
\begin{center}
\caption{Sample of Energy 2D's resulting file}
\label{table:energy2DData}
\begin{tabular}{{@{}llll@{}}}
\toprule
time & T1 & T2 & T3 \\
(s) & (ºC) & (ºC) & (ºC) \\ 
\midrule
100 & -6.96 & -7.01 & -6.97 \\

200 & -4.61 & -5.05 & -4.64 \\ 

300 & -3.24 & -4.05 & -3.29 \\ 
\botrule
\end{tabular}
\end{center}

\end{table} 

\subsubsection{Data Acquisition}
\label{sec:uc2DataAcq}
This module obtains data from the Monitoring component (Sec. \ref{sec:uc1Monitoring}) and our house's computational model (Sec. \ref{sec:uc1CompModel}).

We wrote a Python script named \textit{simulation-data-parser} to transfer the data from the simulation output (Table \ref{table:energy2DData}) to MQTT topics. It maps each of the columns T1, T2, and T3 to a different MQTT topic. Moreover, the Data Acquisition component transfers data from the four MQTT topics, i. e., three from the simulation and the topic \textit{temperature}, to respective Kafka ones with the same names. Same as we did in the first use case, the MQTT broker we used was EMQX, and the MQTT-Kafka mapping we did with the MQTT Connector Source that is available on Confluent Hub. The script does not publish the messages directly on Kafka topics to maintain the MQTT broker as the interface to our data storage. 

\subsubsection{Data Analysis}
\label{sec:uc2DataAna}
After the \textit{simulation-data-parser} has finished, we obtain three topics on Kafka that store the simulation of the heating of the rooms. Moreover, a fourth topic receives real-time information from the physical world. The four streams temporal progressions, having each of its events a temperature, in Celsius, and time information, a value in seconds in the simulation streams, and a time, i. e., hour and minute, in the \textit{temperature} topic.

\subsubsection{Services}
\label{sec:uc2Services}

\begin{figure}
    \centering
    \includegraphics[scale=0.1]{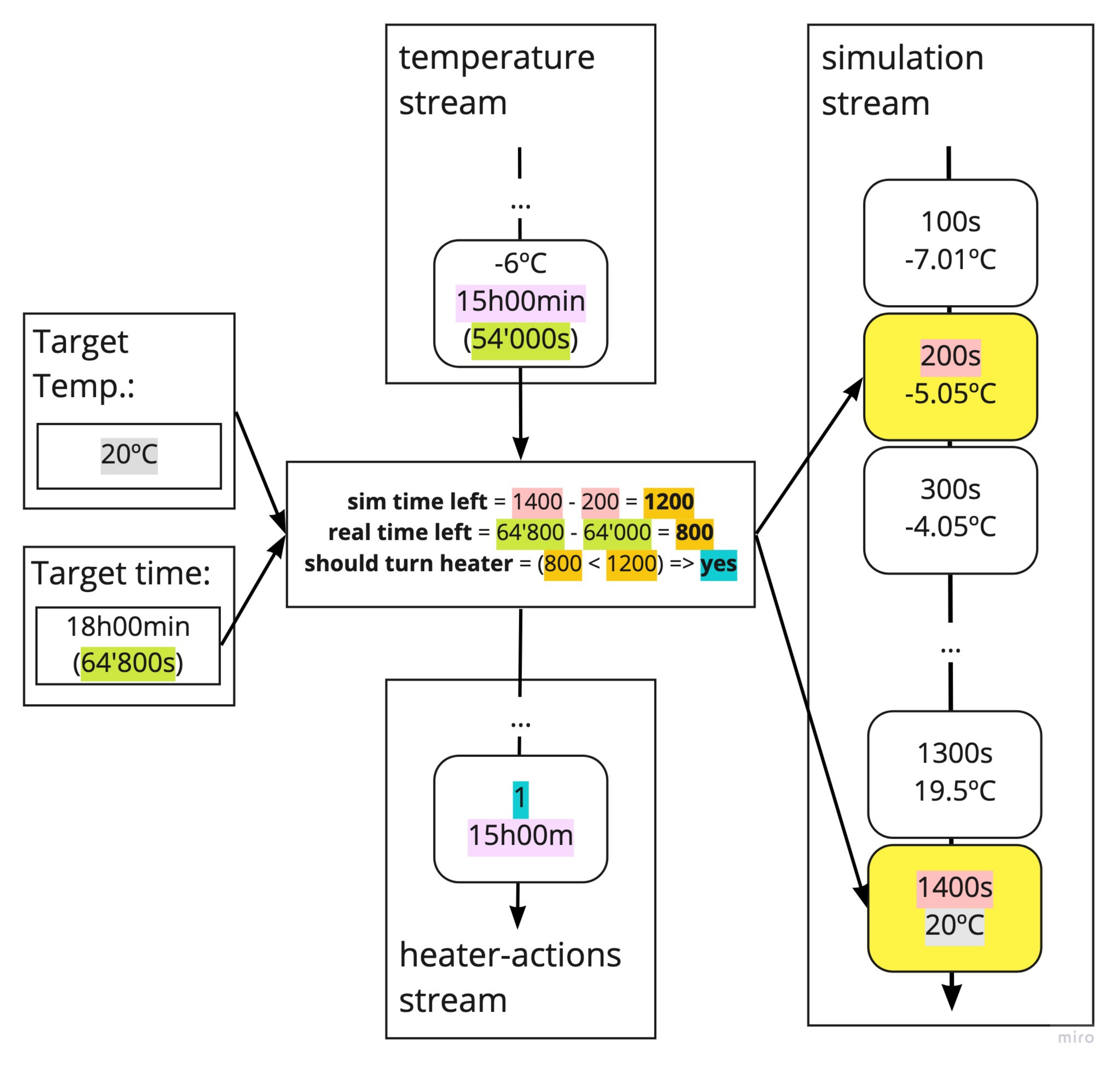}
    \caption{Data Analysis operation}
    \label{fig:dataanalysis}
\end{figure}

In parallel with the \textit{temperature-generator} script, we run another one whose operation we show in Fig. \ref{fig:dataanalysis}. It obtains as input a room name and subscribes to the \textit{temperature} Kafka topic. For every message it receives, the program executes the following steps. 
\begin{enumerate}

\item \label{itm:UC2Item1} Makes the following kSQL query on the simulation stream determined by the input to know when the simulation had a temperature close to the one received. As a result, it obtains a time measurement in seconds;

\begin{verbatim}
SELECT * FROM {stream_name} 
WHERE temperature >= {temperature} 
EMIT CHANGES 
LIMIT 1;
\end{verbatim}

\item \label{itm:UC2Item2} Computes at which moment the simulated space reached 20ºC. If the program is processing the first received message, it executes the previous query with 20ºC as the temperature. Then, it saves the result in a variable for the subsequent messages;

\item \label{itm:UC2Item3} Subtracts item \ref{itm:UC2Item1}'s result from \ref{itm:UC2Item2}'s one, obtaining how long in seconds the simulated room took to heat from the physical room's indoor temperature to the target;

\item \label{itm:UC2Item4} Calculates how much time in seconds is left from the current clock time in the physical room, received on the message, to 18 o'clock; 

\item \label{itm:UC2Item5} Compares the values of items \ref{itm:UC2Item3} and \ref{itm:UC2Item4}. If the latter period is less than or equal to the previous, the script publishes a JSON message in a Kafka topic named \textit{heater-actions}. The construct contains a key indicating this decision as a Boolean and another containing the time included in the event received.

\end{enumerate}

\subsubsection{Connectivity}
\label{sec:uc2Conn}

This module forwards the messages on Kafka's \textit{heater-actions} topic to an MQTT topic with the same name. We again used Kafka's MQTT Connector Sink available on Confluent Hub. 

\subsection{Evaluation}
\label{sec:evaluationUC2}
\begin{figure}
    \centering
    \includegraphics[scale=0.1]{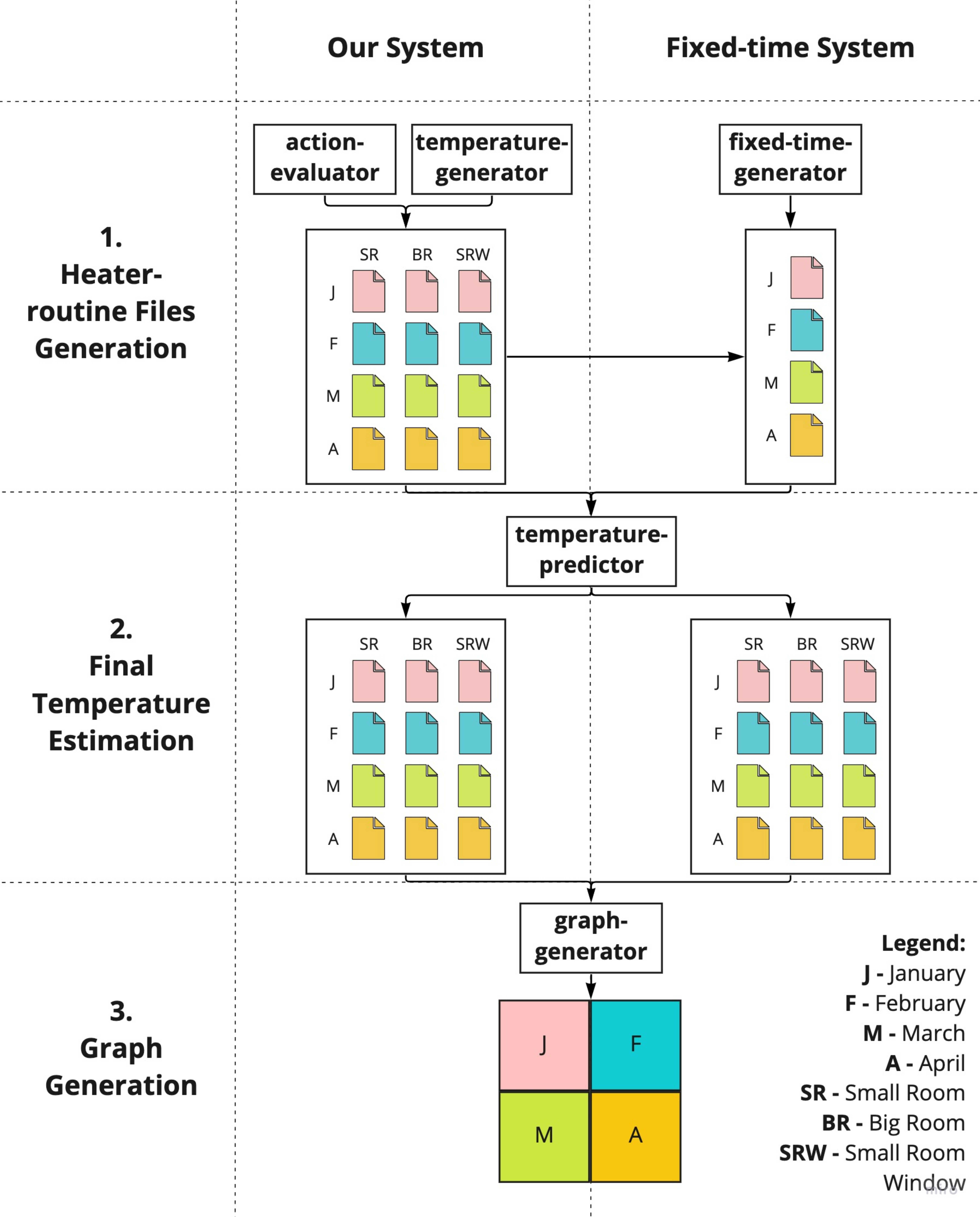}
    \caption{Evaluation process}
    \label{fig:evaluation}
\end{figure}

To evaluate the second use case of the system, we compare it to one that turns on all the heaters every day at the same fixed time, which we further refer to as \textit{fixed-time} system. We analyzed the behavior of the two approaches during the first four months of the year and in three different rooms. Then, we compared them regarding thermal comfort and power consumption. Fig. \ref{fig:evaluation} shows the evaluation's steps explained below.

\paragraph{Heater-routine Files Generation}

To generate the \textit{heater-routine} files describing our system's operation, we run, in parallel, the scripts \textit{temperature-generator} and \textit{action-evaluator}. The first receives a month as a parameter, the second a room name. Thus, we ran twelve iterations, one for each combination between room and month. 

For the fixed-time system, we made a program that generates four files, one for each month. They have the same structure as the \textit{heater-routine} files, but each has the same value in all rows of columns \textit{action-time} and \textit{temperature}. We determine a month's time from the average time among the files generated for our system for the same month, rounded with the frequency level of 15 minutes. We then fetch the temperature in which our city of interest was at the calculated moment based on the \textit{stuttgart-weather} data set. 

\paragraph{Final Temperature Estimation}

A Python script we wrote expects two parameters: a \textit{heater-routine} file and a room name. For each day's \textit{action-time} X, temperature T, it estimates what temperature the given room would have at 6 pm if we turned on its heater at time X and initial temperature T. To accomplish this, it executes the following steps.

\begin{enumerate}
    \item Fetches the timestamp Y at which the simulated room's temperature was the closest to T by executing the following kSQL query on the simulation stream respective to the room. As a result, it obtains an integer representing the number of seconds since the start of the simulation;

\begin{verbatim}

SELECT time FROM {room_name} 
WHERE temperature >= {temperature} 
EMIT CHANGES
LIMIT 1;

\end{verbatim}

    \item Calculates how much time in seconds remains from X until 6 pm;
    
    \item Sums the latter to Y;

    \item Fetches the simulation temperature at the previous item's instant. To do so, it queries the simulation stream with the following operation, obtaining the desired result in Celsius.

\begin{verbatim}

SELECT * from {room_name} 
WHERE measured_time >= {measured_time} 
EMIT CHANGES 
LIMIT 1;

\end{verbatim}
    
\end{enumerate}

The program repeats the procedure for all lines of the input file and outputs the same with an additional column with the simulated final temperature. 

\begin{figure}
\centering
\includegraphics[scale=0.15]{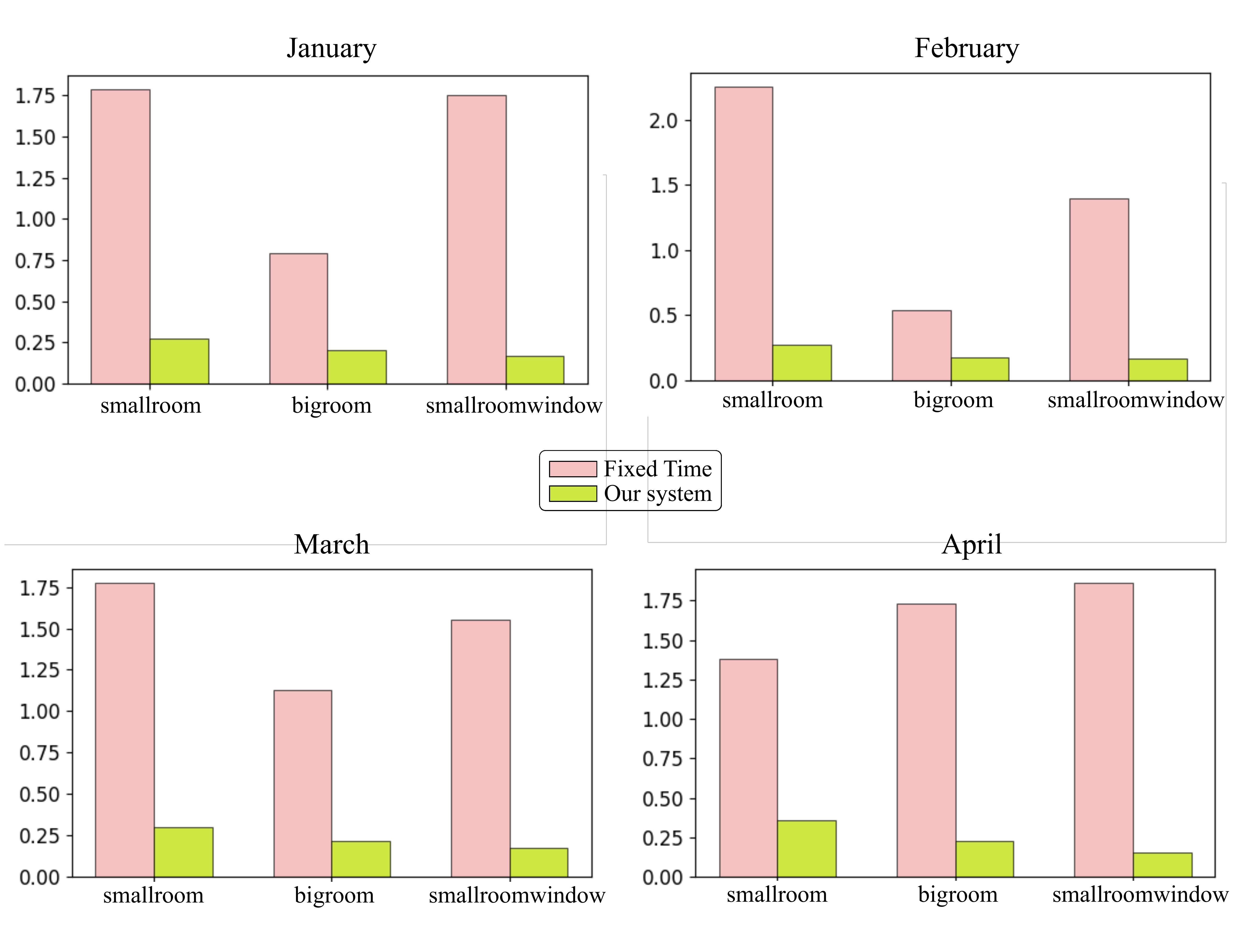}
\caption{Absolute difference between 20ºC and final temperatures}
\label{fig:usecase2results1}
\end{figure}

\paragraph{Graph Generation}
We compare the systems based on their behavior in the 12 scenarios. The first comparison criterion is thermal comfort. We measured the absolute difference between the estimated final temperatures and 20ºC. Then, we made the monthly average of these values for each room. Fig. \ref{fig:usecase2results1} shows the results proving that our system provided better thermal comfort in all scenarios. On average, our system differed from the goal by 0.38 degrees Celsius at 18 o'clock. In turn, the fixed-time system by 2.56. Thus, our approach delivers a thermal comfort approximately 85\% better.

\begin{figure}
\centering
\includegraphics[scale=0.15]{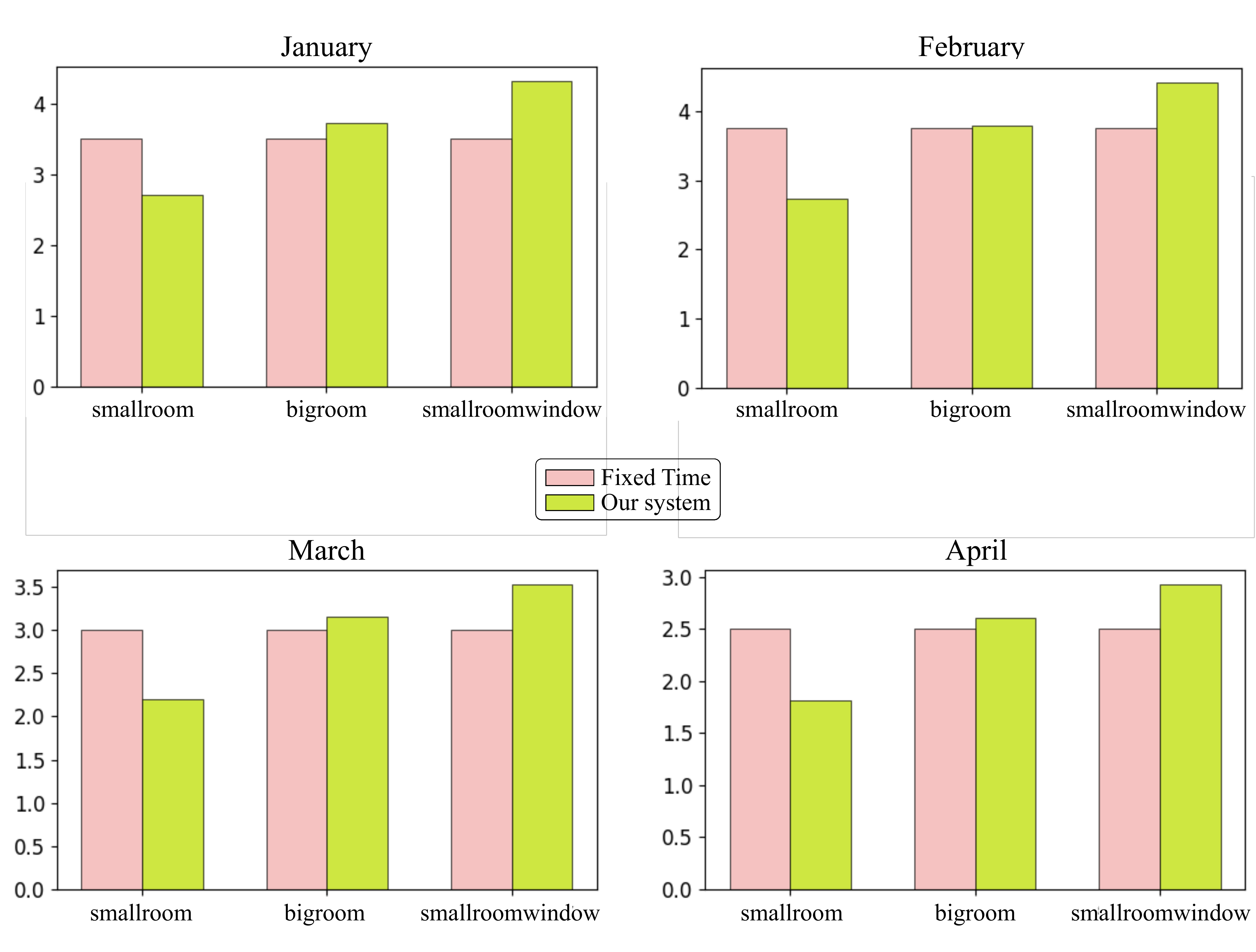}
\caption{Heaters' operation duration in hours}
\label{fig:usecase2results2}
\end{figure}

The second comparison criterion was energy consumption. Fig. \ref{fig:usecase2results2} shows the monthly averages in hours that the heaters worked in each room, considering only the period from when the heater was turned on to 6 pm. Since the \textit{action-times} of the \textit{fixed-time} system are averages of ours, we can see that, per month, both systems consume the same. However, our system better distributes this energy among the rooms.

\section{Conclusion}
\label{sec:conclusion}



This work demonstrated that a home's Digital Twin could contribute with analysis and automation. We proposed an architecture for the system and implemented two use cases that share structure and technologies but have different purposes, contemplating two roles of the DTs. We open-sourced both implementations on GitHub\footnote{\url{https://github.com/Open-Digital-Twin/article-smart-home-use-cases}}.

The first use case focused on the interrogative role of a Digital Twin. It allowed the users to inspect the current state of home appliances regarding their power consumption and to analyze the average of the same property during a custom period. We presented the data with 3D replicas of the machinery colored by a color-coded gradient. 

The second use case focused on the DT's predictive role. Based on heating simulations, it could determine the best time to turn on power sources in different rooms so that their indoor temperature equals a target value at a particular time. We compared our approach to one that turns all the heaters on every day at the same time and could improve the house's thermal comfort by 85\% with the same energy consumption.

Our implementations' main limitations are twofold: the absence of an actual physical home and the possible imprecise heating simulation in the second use case, which we made in 2D software disregarding factors that contribute to a room's temperature, such as furniture and inhabitants. Future work could implement our project in a real house and use preciser software to simulate a house's heating, such as EnergyPlus\footnote{\url{https://energyplus.net/} Visited on 28. Mar. 2022}, which considers the characteristics of the building and the climate in its region. Moreover, as we did not present a scalability test in this work, another expansion could analyze the feasibility of expanding our systems to larger numbers of residences and sensors.



\bibliography{sn-bibliography}
\end{document}